\documentclass[
  aps,
  prd,
  onecolumn,
  12pt,
  notitlepage,
  superscriptaddress,
  nofootinbib,
  nolongbibliography,
  floatfix,
  amsmath,
  amssymb
]{revtex4-2}

\usepackage{graphicx}
\usepackage{tikz}
\usetikzlibrary{decorations.pathmorphing,decorations.markings,arrows.meta}
\usepackage{dcolumn}
\usepackage{bm,comment}
\usepackage{placeins}
\usepackage[colorlinks=true,allcolors=blue]{hyperref}

\newcommand{\Lag}{\mathcal{L}}

\begin{document}


\title{Vector Dark Matter from Gravitational Quantum Field Theory}

\author{Qiu-Xi Zhang}
\affiliation{University of Chinese Academy of Sciences (UCAS), Beijing 100049, China}
\affiliation{International Center for Theoretical Physics Asia-Pacific, Beijing, China}

\author{Yong Tang}
\affiliation{University of Chinese Academy of Sciences (UCAS), Beijing 100049, China}
\affiliation{International Center for Theoretical Physics Asia-Pacific, Beijing, China}
\affiliation{School of Fundamental Physics and Mathematical Sciences, \\Hangzhou Institute for Advanced Study, UCAS, Hangzhou 310024, China}

\author{Yue-Liang Wu}
\affiliation{University of Chinese Academy of Sciences (UCAS), Beijing 100049, China}
\affiliation{International Center for Theoretical Physics Asia-Pacific, Beijing, China}
\affiliation{School of Fundamental Physics and Mathematical Sciences, \\Hangzhou Institute for Advanced Study, UCAS, Hangzhou 310024, China}
\affiliation{Institute of Theoretical Physics, Chinese Academy of Sciences, Beijing 100190, China}
\affiliation{Taiji Laboratory for GW Universe (Beijing/Hangzhou), University of Chinese Academy of Sciences, Beijing 100049, China}

\date{\today}

\begin{abstract}

Gravitational evidence firmly establishes the existence of dark matter (DM), yet its non-gravitational interactions and fundamental nature remain unresolved. In this work, we demonstrate that a chirality boost-spin gauge field, which emerges naturally in the General Standard Model formulated within gravitational quantum field theory, offers a vector DM candidate. We construct the corresponding low-energy effective field theory for this DM with axial-vector couplings to Standard Model fermions and systematically investigate its signatures in direct detection, indirect detection, and collider experiments. Comparing our predictions against current experimental data places stringent constraints on the DM parameter space, with the viable regions within the further reach of future searches.
\end{abstract}

\maketitle
\clearpage

\section{Introduction}
\label{sec:introduction}

The evidence for dark matter (DM) is supported by a wide range of cosmological and astrophysical observations~\cite{CloweBulletCluster2006, Planck2018}, including cosmic microwave background, large-scale structure, bullet cluster, and galactic rotation curve. Nevertheless, the physical nature of DM, such as the mass, spin and possible nongravitational interactions, remain unknown~\cite{BertoneParticleDM2005, FreeseStatusDM2017}. At present, there is a general consensus that DM must participate in gravitational interactions, and it is conceivable that DM could have its origin within the gravitational sector itself.

A promising framework in this direction is Gravitational Quantum Field Theory (GQFT) \cite{WuGQFT2016, WuGravidynamics2023, WuGravitizationZeroEMT2025,Gao2024LinearDynamics, Gao2025GravitationalWaves, Xu2025Polarization,Xu2026ScalarMode}, which aims to unify general relativity with quantum field theory by gauging the intrinsic spin symmetry of elementary fermions. When this formalism is extended to incorporate the full matter content of the Standard Model (SM), including its established gauge symmetries $\mathrm{U}(1)\times\mathrm{SU}_L(2)\times\mathrm{SU}_c(3)$ for the electroweak and strong interactions, it leads to the General Standard Model (GSM)~\cite{WuGSMFoundation2025, WuGSMDarkSide2025}. A distinctive feature of the GSM is the emergence of the conformal inhomogeneous spin gauge symmetry, denoted as $\mathrm{WS}_c(1,3)=\mathrm{SP}(1,3)\rtimes W^{1,3}\rtimes\mathrm{SP}_c(1,1)$ for chiral leptons and quarks, where $\mathrm{SP}(1,3)$ denotes the rotation-like spin gauge symmetry, $W^{1,3}$ the translation-like chirality boost-spin gauge symmetry, and $\mathrm{SP}_c(1,1)$ the scaling-like chiral conformal-spin gauge symmetry. This chiral-related gauge symmetry arises naturally as a consequence of the maximal parity violation observed in weak interactions.

Within this framework, the chirality boost–spin gauge field $\mathcal{W}_{\mu}^{a}(x)$ naturally emerges as a massive vector boson stabilized by a discrete $Z_2$ symmetry, and therefore constitutes a compelling DM candidate, referred to as the dark graviton. Its interactions with SM particles are mediated by a heavy spin gauge field  $\mathcal{A}_{\mu}^{ab}(x)$ associated with the SP(1,3) sector. One of the most attractive features of the GSM is that the interaction structure of the DM is constrained by the underlying gauge symmetry, endowing the framework with considerable predictive power. Nevertheless, in close analogy to the massive electroweak gauge bosons of the SM, the exact values of the masses and coupling constants remain theoretically unconstrained and must be determined through experimental measurements. Various experimental searches for DM provide stringent constraints on these parameters.

In this paper, we present the low-energy effective field theory for the massive vector DM, and investigate its phenomenology in direct detection, indirect detection, and collider searches. We show the vector DM interacts with SM fermions through an axial-vector portal, calculate the relevant scattering and annihilation cross sections and compare with the current experimental results.  In the absence of any confirmed DM signal, we compile the current experimental limits into exclusion bounds on the DM mass and coupling. The remaining parameter regions will provide valuable guidance for future experimental efforts aimed at detecting the vector DM predicted by the GSM within the GQFT framework.

The paper is organized as follows.  In Section.~\ref{sec:model} we briefly review the theoretical formalism for GQFT
and derive the low-energy effective operators for the coupling of vector DM to SM fermions. In Sec.~\ref{sec:phenomenology}, we discuss the relevant phenomenology, calculate the corresponding scattering or annihilation cross sections for direct detection, collider, and indirect detection searches, and present the combined constraints on the mass and couplings of DM. Finally, we conclude in Sec.~\ref{sec:conclusions}. 

\section{Theoretical Formalism}
\label{sec:model}

The DM candidate considered in this work arises from the General Standard Model (GSM), which is formulated within the framework of gravitational quantum field theory (GQFT)~\cite{WuGQFT2016, WuGravidynamics2023, WuGravitizationZeroEMT2025, WuGSMFoundation2025, WuGSMDarkSide2025}.
In this framework, the laws of nature are governed by the intrinsic properties of the
basic constituents of matter, including the gravitational interactions, and a strict distinction is made between intrinsic and external symmetries. In conventional quantum field theory (QFT), the Lorentz invariance of the electron spinor field requires that the global Lorentz transformation of coordinates, governed by the symmetry group SO(1,3), coincides with the global spin transformation of the spinor field under the spin group SP(1,3) in the spinor representation. Thus, SO(1,3) and SP(1,3) form an associated symmetry.

In GQFT, however, the global Lorentz symmetry
$\mathrm{SO}(1,3)$ of coordinates is defined on Minkowski spacetime (Greek indices $\mu,\nu$),
while the intrinsic spin symmetry $\mathrm{SP}(1,3)$ is defined in the spinor
Hilbert space (Latin indices $a,b$) and is localized in accordance with the gauge
principle. These two symmetries combine to form the joint symmetry
$\mathrm{SO}(1,3)\Join\mathrm{SP}(1,3)$: 
over each point of the flat Minkowski base sits an intrinsic spin
space, the fiber in which the spinors and their local $\mathrm{SP}(1,3)$
rotations act, and the two are connected by the gravigauge field
$\chi_\mu{}^a(x)$, with dual $\hat\chi_a{}^\mu$. This field transforms homogeneously
under both symmetries~\cite{WuGravitizationZeroEMT2025, WuGQFT2016, WuGravidynamics2023} and relates to the standard metric field through $g_{\mu\nu}=\chi_\mu{}^a\chi_\nu{}^b\eta_{ab}$.

By promoting the spin symmetry of lepton and quark spinor fields to a local gauge symmetry, the SM within the framework of QFT is extended into GQFT. Its explicit form for the lepton and quark sectors is given by,
\begin{align}
  \Lag_f^{\mathrm{GQFT}}
  &=
  \chi\sum_i\left\{
  \frac{i}{2}\,\hat\chi_a{}^{\mu}
  \left[
    \bar\psi_i\gamma^a D_\mu\psi_i
    -
    (D_\mu\bar\psi_i)\gamma^a\psi_i
  \right]
  -
  m_i\bar\psi_i\psi_i
  \right\},
  \nonumber\\
  iD_\mu\psi_i
  &=
  \left(
    iD_\mu^{\mathrm{SM}}
    +\frac{g_4}{2}\,\mathcal A_\mu^{ab}\Sigma_{ab}
  \right)\psi_i,
  \qquad
  \Sigma_{ab}=\frac{i}{4}[\gamma_a,\gamma_b],
  \label{eq:gqft-dirac}
\end{align}
with the sum running over the SM leptons and quarks ($\psi_i$). Here $\chi=\det\chi_\mu{}^a$, the covariant derivative is $iD_\mu\bar\psi_i=iD_\mu^{\mathrm{SM}}\bar\psi_i - \bar\psi_i\frac{g_4}{2}\mathcal A_\mu^{ab}\Sigma_{ab} $, and the masses $m_i$ are generated by the Yukawa sector. The gauge invariance principle introduces the spin-gauge
field $\mathcal A_\mu^{ab}=-\mathcal A_\mu^{ba}$, valued in the generators
$\Sigma_{ab}$ with coupling $g_4$. This field enters the covariant derivative
alongside the SM gauge fields and acts as the heavy mediator at lower energies.

In the flat background limit, ignoring gravitational interactions, we have $\chi_\mu{}^a \to \delta_\mu{}^a$, $\chi\to1$, and
Eq.~\eqref{eq:gqft-dirac} reduces to the standard QFT Lagrangian
\begin{equation}
  \Lag_f
  =
  \frac{i}{2}\sum_i
  \left[
    \bar\psi_i\gamma^\mu D_\mu\psi_i
    -
    (D_\mu\bar\psi_i)\gamma^\mu\psi_i
  \right]
  -
  \sum_i m_i\bar\psi_i\psi_i ,
  \label{eq:fermion-kinetic-symmetric}
\end{equation}
with $\gamma^\mu\equiv\gamma^a\delta_a{}^{\mu}$: the SM Dirac sector thus acquires
a single new term, the spin-gauge mediator $\mathcal{A}_{\mu}^{ab}$ in covariant derivative $D_\mu$.

Recognizing that all SM leptons and quarks are described by Weyl fields, owing to maximal parity violation in the weak interactions, the GSM~\cite{WuGSMFoundation2025,WuGSMDarkSide2025} retains the SM gauge
group while extending the internal spin gauge symmetry SP(1,3) of leptons and quarks to the conformal inhomogeneous spin gauge symmetry,
\[
  \mathrm{WS}_c(1,3)
  =
  \mathrm{SP}(1,3)\rtimes W^{1,3}\rtimes\mathrm{SP}_c(1,1).
\]
Under $\mathrm{WS}_c(1,3)$, the SM fermions assemble into the equivalent chiral multiplets $\Psi_-=(\Psi_L,\Psi_R)^T$ and $\Psi_+=(\Psi_R,\Psi_L)^T$, collectively denoted by $\Psi_s$, $s=\pm$. The translation-like chirality boost-spin gauge symmetry $W^{1,3}$ gives rise to the massive vector gauge field $\mathcal W_\mu^a$, which does not directly couple to the Weyl fermions due to chirality and is naturally stabilized by an emergent discrete symmetry $Z_2$. This vector gauge boson serves as the dark-matter candidate considered in this work. 

Retaining only the connections relevant to this work, the fermion sector reads (families summed; Yukawa and $\mathrm{SU}_C(3)$ terms omitted)
\begin{align}
  \Lag^{(f)}_{\mathrm{GSM}}
  &=
  \chi\,\frac12\sum_{s=\mp}\sum_{\Psi=l,q}
  \left[
    \frac12\,\bar\Psi_s\,\Sigma_s^{c}\,\hat\chi_c{}^{\mu}\,
    i\mathcal D_\mu\Psi_s
    +\mathrm{H.c.}
  \right],
  \nonumber\\
  i\mathcal D_\mu\Psi_s
  &=
  \left[
    i\partial_\mu
    +g'B_\mu\Sigma_{Ys}^{(\Psi)}
    +g\,W_\mu^{i}\Sigma_{Ls}^{i}
    +\frac{g_4}{2}
    \left(
      \mathcal A_\mu^{ab}\Sigma_{ab}
      +\mathcal W_\mu^{a}\Sigma_{a s}
    \right)
  \right]\Psi_s .
  \label{eq:gsm-fermion}
\end{align}
In the covariant derivative above, the displayed SM connections include the gauge fields $B_\mu$ and $W_\mu^i$ corresponding to the electroweak gauge symmetries $\mathrm{U}_Y(1)\times\mathrm{SU}_L(2)$, with generators
$\Sigma_{Ys}^{(\Psi)}$ and $\Sigma_{Ls}^i$, respectively.
The additional gauge fields \(\mathcal A_\mu^{ab}\) and \(\mathcal W_\mu^a\) correspond to the \(\mathrm{SP}(1,3)\) and \(W^{1,3}\) gauge symmetries of \(\mathrm{WS}_c(1,3)\), respectively, with generators \(\Sigma_{ab}=i[\Gamma_a,\Gamma_b]/4\) and \(\Sigma_{as}=\Gamma_a\Gamma_s\). Here $\Gamma_{s}=(1+s\gamma_{9})/2$, with $s=\pm$, and $\gamma_9=\sigma_3\otimes\sigma_0\otimes\gamma_5$. These generators share a 16-dimensional Clifford-algebra representation, with the SM quantum numbers fixed by chiral projectors~\cite{WuGSMFoundation2025,WuGSMDarkSide2025}.
$\Sigma_s^c\hat\chi_c{}^\mu$ plays the role of the Dirac matrix in the fermion kinetic term and reduces to the usual $\gamma^\mu$ structure when gravitational interactions are neglected, i.e., when $\hat\chi_c{}^\mu\to\delta_c{}^\mu$.

The chirality boost-spin generators
obey the Poincar\'e-like relations
\[
  [\Sigma^{ab},\Sigma_s^c]
  =
  i\left(
    \Sigma_s^a\eta^{bc}
    -
    \Sigma_s^b\eta^{ac}
  \right),
  \qquad
  [\Sigma_s^a,\Sigma_s^b]=0.
\]
Notably, in this chiral representation, the stronger identity
$\Sigma_s^a\Sigma_s^b=0$ holds due to chirality property. Consequently, the
$\mathcal W_\mu^a\Sigma_{as}$ term in Eq.~\eqref{eq:gsm-fermion} decouples via,
\[
  \bar\Psi_s\Sigma_s^c\hat\chi_c{}^\mu
  \mathcal W_\mu^a\Sigma_{as}\Psi_s
  =
  0.
\]
Together with the even-$\mathcal W$ bosonic sector, this feature endows the low-energy
theory with an accidental $Z_2^W$ symmetry: $\mathcal W_\mu^a\to-\mathcal W_\mu^a$, with all other fields even.
This symmetry stabilizes $\mathcal W_\mu^a$, whose interactions with SM fermions are mediated via the spin gauge boson $\mathcal A_\mu^{ab}$.

We take this mediator to be heavy
(see Ref.~\cite{Huang:2022jpl} for related collider constraints).
Ignoring other interactions and symmetry-breaking sectors in the GSM,
we retain only $\mathcal W_\mu^a$ and $\mathcal A_\mu^{ab}$
to derive the low-energy effective interactions.
The tree-level dynamics then depends on the masses $m_{\mathcal W}$,
$M_{\mathcal A}$, and the coupling $g_4$, and the relevant Lagrangian reads 
\begin{align}
  \Lag_{\mathrm{rel}}
  &=
  \Lag_f
  -
  \frac{1}{4}
  \mathcal F_{\mu\nu}^{ab}\mathcal F^{\mu\nu}{}_{ab}
  +
  \frac{1}{4}
  \mathcal W_{\mu\nu}^{a}\mathcal W^{\mu\nu}{}_{a}
  +
  \frac{1}{2}M_{\mathcal A}^2
  \mathcal A_\mu^{ab}\mathcal A^{\mu}{}_{ab}
  -
  \frac{1}{2}m_{\mathcal W}^2
  \mathcal W_\mu^a\mathcal W^{\mu}{}_{a},
  \label{eq:relevant-lagrangian}
\end{align}
with the SM gauge and Higgs terms left implicit.  The corresponding field
strengths are
\begin{align}
  \mathcal F_{\mu\nu}^{ab}
  &=
  \partial_\mu\mathcal A_\nu^{ab}
  -\partial_\nu\mathcal A_\mu^{ab}
  +g_4(\mathcal A_{\mu}{}^a{}_{c}\mathcal A_\nu^{cb}
       -\mathcal A_{\nu}{}^a{}_{c}\mathcal A_\mu^{cb}),
  \nonumber\\
  \mathcal W_{\mu\nu}^{a}
  &=
  D_\mu\mathcal W_\nu^a-D_\nu\mathcal W_\mu^a,
  \qquad
  D_\mu\mathcal W_\nu^a
  =
  \partial_\mu\mathcal W_\nu^a
  +g_4\mathcal A_{\mu}{}^a{}_{b}\mathcal W_\nu^b .
  \label{eq:field-strengths}
\end{align}

The fermionic part $\Lag_f$ is readily obtained from the above action. Hermiticity converts the term linear in the spin-gauge field into the anticommutator $\{\gamma^c,\Sigma_{ab}\}$, which defines the SM spin current,
\begin{equation}
  \Lag_{\mathcal A\psi}
  =
  \mathcal A_\mu^{ab}J_{f\,ab}^\mu,
  \qquad
  J_{f\,ab}^\mu
  =
  \frac{g_4}{4}
  \sum_i \hat{\chi}_c^{\mu}
  \bar\psi_i\{\gamma^c,\Sigma_{ab}\}\psi_i ,
  \label{eq:spin-current}
\end{equation}
which couples to the heavy mediator $\mathcal A_\mu^{ab}$.  The
$\mathcal W_\mu^a$ field couples to the same mediator: expanding the kinetic term of $\mathcal W_\mu^a$ to linear order in $\mathcal A_\mu^{ab}$ yields a second current,
\begin{equation}
  \Lag_{\mathcal A\mathcal W\mathcal W}
  =
  \mathcal A_{\mu}^{ab}J_{\mathcal W\,ab}^{\mu},
  \qquad
  J_{\mathcal W\,ab}^{\mu}
  =
  g_4\mathcal W^{\mu\nu}_{[a}\mathcal W_{\nu b]},
  \label{eq:w-current}
\end{equation}
with $\mathcal W^{\mu\nu}_a \equiv \eta_{ab}\hat{\chi}^{\mu\mu'} \hat{\chi}^{\nu\nu'} \mathcal W_{\mu'\nu'}^b$ evaluated at its abelian order, $\partial_\mu\mathcal W_\nu^a-\partial_\nu\mathcal W_\mu^a$, since Eq.~\eqref{eq:w-current} is already linear in $\mathcal A_\mu^{ab}$. Here $\mathcal W^{\mu\nu}_{[a}\mathcal W_{\nu b]}\equiv \tfrac12(\mathcal W^{\mu\nu}_{a}\mathcal W_{\nu b}
-\mathcal W^{\mu\nu}_{b}\mathcal W_{\nu a})$ denotes the weight-$\tfrac12$ antisymmetrization in the internal indices.

At energies well below $M_{\mathcal A}$, the mediator can be
integrated out~\cite{BuchmuellerBeyondEFT2014, AbercrombieDMForum2019}. 
We adopt the metric signature $(+,-,-,-)$ and define $\gamma_5=i\gamma^0\gamma^1\gamma^2\gamma^3$. Using
the Dirac identity
\begin{equation}
  \{\gamma_c,\Sigma_{ab}\}
  =
  \epsilon_{c ab d}\,
  \gamma^d\gamma_5,
  \qquad
  \epsilon_{0123}=+1.
  \label{eq:gamma-identity}
\end{equation}
The internal indices \(a,b\) are associated with the spin-gauge field, with the explicit contractions given in Appendix~\ref{app:frame-direct-detection}. The cross term between the two currents then gives rise to the dimension-six axial-vector portal,
\begin{align}
  \Lag_{\mathrm{eff}}^{(\mathcal W-\psi)}
  &=
  -\frac{1}{M_{\mathcal A}^2}J_{f\,\mu ab}J_{\mathcal W}^{\mu ab}
  =
  -\frac{g_4^2}{4M_{\mathcal A}^2}\,
  \epsilon_{c ab d}\,
  \mathcal W^{\nu b}
  \left(\partial_\mu\mathcal W_\nu^{\,a}
        -\partial_\nu\mathcal W_{\mu}^a\right)
  \hat{\chi}^{\mu c} \sum_i\bar\psi_i\gamma^d\gamma_5\psi_i .
  \label{eq:dimension-six}
\end{align}
The elastic scattering amplitudes are governed by $g_4^2/M_{\mathcal A}^2$, so that experimental limits on scattering impose bounds on $M_{\mathcal A}/g_4$, or equivalently $M_{\mathcal A}$ for fixed $g_4$, under the conditions $|q^2|\ll M_{\mathcal A}^2$ and $g_4^2/(4\pi)\lesssim1$. In the present analysis for low-energy effective action, gravitational effects are negligible. Therefore we can impose $\chi_{\mu}^c \to \delta_{\mu}^c$.

For the following phenomenological analysis, we assume that $\mathcal W$ accounts for all of the dark matter.
Its abundance may be set either through thermal annihilation or via other cosmological scenarios in the early universe~\cite{TangWuPureGravDM2016, TangWuWeylDM2020, Tang:2020ovf, Kolb:2023ydq}. The precise thermal history would depend on its full set of interactions with other particles in the theory, in addition to the SM fermions.

\section{Phenomenology}
\label{sec:phenomenology}
Equipped with the effective operator in
Eq.~\eqref{eq:dimension-six}, we now quantitatively investigate the
phenomenology of the vector DM. We derive its elastic-scattering and
annihilation cross sections and examine the corresponding signatures in
direct-detection, indirect-detection, and collider searches, including
LEP-2 fermion-pair production and the mono-jet channel. We compare these
predictions with existing experimental limits and present the resulting
constraints, together with the expected PandaX-xT sensitivity, in the
$(m_{\mathcal W},M_{\mathcal A}/g_4)$ parameter space.

\subsection{Direct detection}
\label{subsec:direct-detection}

In the nonrelativistic limit, the operator in
Eq.~\eqref{eq:dimension-six} reduces to a spin-dependent contact interaction
between the $\mathcal W$ and fermion spins. We keep the leading
velocity-independent term. Details of the calculation are provided in Appendix~\ref{app:frame-direct-detection}.

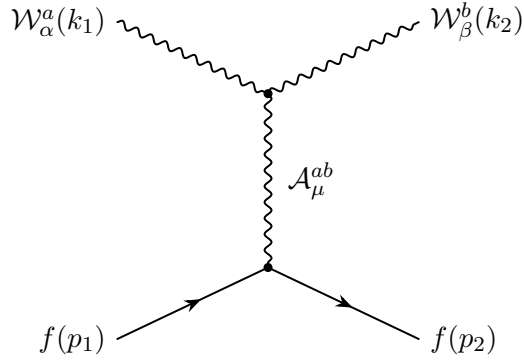
\begin{figure}[t]
  \centering
  \begin{tikzpicture}[line width=0.75pt,
    boson/.style={decorate, decoration={snake, amplitude=1.3pt, segment length=6pt}},
    fer/.style={postaction={decorate},
      decoration={markings, mark=at position 0.56 with {\arrow{Stealth}}}}]
    \coordinate (vW) at (0,1.15);
    \coordinate (vf) at (0,-1.15);
    \draw[boson] (-2,2.1) -- (vW);
    \draw[boson] (vW) -- (2,2.1);
    \draw[fer] (-2,-2.1) -- (vf);
    \draw[fer] (vf) -- (2,-2.1);
    \draw[boson] (vW) -- (vf);
    \fill (vW) circle (1.7pt);
    \fill (vf) circle (1.7pt);
    \node[left] at (-2,2.1) {$\mathcal W_\alpha^a(k_1)$};
    \node[right] at (2,2.1) {$\mathcal W_\beta^b(k_2)$};
    \node[left] at (-2,-2.1) {$f(p_1)$};
    \node[right] at (2,-2.1) {$f(p_2)$};
    \node[right] at (0.12,0) {$\mathcal A_\mu^{ab}$};
  \end{tikzpicture}
  \caption{Tree-level exchange of the heavy spin-gauge mediator
  $\mathcal A_\mu^{ab}$ in the elastic process
  $\mathcal W_\alpha^a(k_1)f(p_1)\to\mathcal W_\beta^b(k_2)f(p_2)$, where $f$
  denotes a SM fermion. For $|q^2|\ll M_{\mathcal A}^2$, the exchange reduces
  to the contact operator in Eq.~\eqref{eq:dimension-six}.}
  \label{fig:feynman-elastic}
\end{figure}

Elastic scattering off a SM fermion $f$, corresponding to one of the $\psi_i$
in Eq.~\eqref{eq:fermion-kinetic-symmetric}, proceeds via the exchange depicted
in Fig.~\ref{fig:feynman-elastic}. The external polarization vectors carry the
coordinate vector index of the massive $\mathcal W_\mu^a$ field. Using the
Dirac identity in Eq.~\eqref{eq:gamma-identity}, the tree-level amplitude takes
the axial form
\begin{equation}
  \mathcal M_{ab}
  =
  -i\frac{g_4^2}{4M_{\mathcal A}^2}
  \widetilde V^{\mu\alpha\beta}
  \varepsilon_{1\alpha}(k_1)
  \varepsilon^*_{2\beta}(k_2)
  \epsilon_{\mu ab\lambda}\,
  \bar u(p_2)\gamma^\lambda\gamma_5u(p_1),
  \label{eq:scattering-amplitude}
\end{equation}
where the crossed $\mathcal A\mathcal W\mathcal W$ vertex tensor is
\begin{equation}
  \widetilde V^{\mu\alpha\beta}(k_1,k_2)
  =
  (k_1+k_2)^\mu\eta^{\alpha\beta}
  -k_1^\beta\eta^{\mu\alpha}
  -k_2^\alpha\eta^{\mu\beta}.
  \label{eq:scattering-v-tensor}
\end{equation}
Using the initial-state averaging factor
$\tfrac12\times\tfrac13\times\tfrac13$ for the fermion spin and the
polarization and internal states of $\mathcal W$, the leading
velocity-independent cross section in the nonrelativistic limit is
\begin{equation}
  \sigma_f^{(0)}
  =
  \frac{g_4^4\mu_f^2}{8\pi M_{\mathcal A}^4},
  \qquad
  \mu_f=
  \frac{m_{\mathcal W}m_f}{m_{\mathcal W}+m_f}.
  \label{eq:fermion-cross-section}
\end{equation}

For electrons, the same leading amplitude takes the spin-dependent
nonrelativistic form
\begin{equation}
  \mathcal M_{\mathrm{NR}}^{(e)}
  =
  c_{\mathrm{SD}}^{(e)}\,
  \mathbf S_{\mathcal W}\cdot\mathbf S_e,
  \qquad
  c_{\mathrm{SD}}^{(e)}
  =
  \frac{g_4^2}{2M_{\mathcal A}^2}.
  \label{eq:electron-sd-operator}
\end{equation}
Here $\mathbf S_{\mathcal W}$ acts on the compact internal triplet of
$\mathcal W$, whereas $\mathbf S_e$ acts on the electron spin. In the basis used
here,
\[
  \left(S_{\mathcal W}^{i}\right)^{a}{}_{b}
  =-i\epsilon^{ia}{}_{b},
  \qquad
  S_e^{i}=\frac{\sigma^{i}}{2},
  \qquad
  \mathbf S_{\mathcal W}\cdot\mathbf S_e
  =\sum_{i=1}^{3}S_{\mathcal W}^{i}S_e^{i},
\]
with $a,b,i=1,2,3$. We use
$\sigma_e^{\mathrm{LO}}\equiv\sigma_f^{(0)}|_{f=e}$ as the free-electron
reference cross section. The electron-recoil limits are taken from the
PandaX-4T analysis for spin-one dark matter in
Ref.~\cite{LiangElectronEFT2024}, the SENSEI single-electron
result~\cite{SENSEISingleElectron2025}, the DAMIC-M
result~\cite{DAMICMHiddenSector2025}, and the XENONnT $7.83~{\rm tonne\,yr}$
ionization-only result~\cite{XENONnTLightDM2026}. At leading
nonrelativistic order and within the independent-electron approximation,
we translate these limits into bounds on $M_{\mathcal A}/g_4$,
restricting each limit to its published mass range.

At leading order, we express the universal quark axial current in
Eq.~\eqref{eq:dimension-six} in terms of nucleon axial matrix elements as
\begin{equation}
  \sum_q\bar q\gamma^\lambda\gamma_5 q
  \to
  \sum_{N=p,n}a_N\,\bar N\gamma^\lambda\gamma_5N,
  \qquad
  a_N=\sum_{q=u,d,s}\Delta q^{(N)} ,
  \label{eq:nucleon-axial-matching}
\end{equation}
where $\Delta q^{(N)}$ are the usual nucleon spin fractions. This yields the
tree-level spin-dependent nucleon cross section,
\begin{equation}
  \sigma_N^{\mathrm{SD}}
  =
  \frac{g_4^4\mu_N^2}{8\pi M_{\mathcal A}^4}
  |a_N|^2,
  \qquad
  \mu_N=
  \frac{m_{\mathcal W}m_N}{m_{\mathcal W}+m_N}.
  \label{eq:nucleon-sd-cross-section}
\end{equation}
With the adopted nucleon axial charges, the universal quark couplings give
$a_p=a_n=0.330$~\cite{HoferichterNuclearResponses2020}.
We approximately convert the published limits for coupling to neutrons
alone to equal proton and neutron couplings by comparing the predicted
recoil rates, including the momentum dependence of the nuclear
responses~\cite{HoferichterNuclearResponses2020,KlosSD2013}.
The nuclear-recoil constraint at each mass is the strongest of the LZ,
PandaX-4T, and XENONnT limits~\cite{LZ2024,LZLowMass2026,PandaX2025,PandaXErratum2026,XENONnTLightDM2026},
while the projected $200~{\rm tonne\,yr}$ PandaX-xT sensitivity illustrates
the future reach~\cite{PandaXxT2025}.
Together with the electron-recoil results, they give the limits in
Fig.~\ref{fig:direct-detection-envelope}: electron scattering extends the
sensitivity to sub-GeV masses, while nuclear recoils constrain heavier
dark matter.
\FloatBarrier

\subsection{Indirect detection}
\label{subsec:indirect-detection}

The same effective operator mediates the annihilation process
\(\mathcal W\mathcal W\to f\bar f\). The squared amplitude is proportional to
\(s^2-m_{\mathcal W}^2s-12m_{\mathcal W}^4\) and vanishes at threshold
\(s=4m_{\mathcal W}^2\), so the \(s\)-wave contribution is absent. For
\(m_{\mathcal W}>m_f\), the leading nonrelativistic cross section is
\begin{equation}
  \sigma(\mathcal W\mathcal W\to f\bar f)v_{\mathrm{rel}}
  =
  b_f v_{\mathrm{rel}}^2
  +\mathcal O(v_{\mathrm{rel}}^4),
  \qquad
  b_f
  =
  \frac{7N_c^f}{2592\pi}
  \frac{g_4^4m_f^2}{M_{\mathcal A}^4}
  \left(1-\frac{m_f^2}{m_{\mathcal W}^2}\right)^{1/2}.
  \label{eq:annihilation-pwave-cross-section}
\end{equation}
Here \(N_c^f=1\) for leptons and \(3\) for quarks. The displayed leading
term exhibits both \(p\)-wave and helicity suppression. Away from the corresponding thresholds,
representative coefficients normalized to \(M_{\mathcal A}/g_4=1~{\rm TeV}\) are
\begin{align*}
  b_{b\bar b}&\simeq5.3\times10^{-31}
  \left(\frac{1~{\rm TeV}}{M_{\mathcal A}/g_4}\right)^4
  ~{\rm cm^3\,s^{-1}},\\
  b_{t\bar t}&\simeq9.0\times10^{-28}
  \left(\frac{1~{\rm TeV}}{M_{\mathcal A}/g_4}\right)^4
  ~{\rm cm^3\,s^{-1}}.
\end{align*}
For Galactic halo velocities, \(v_{\rm rel}^2/c^2\sim10^{-6}\), strongly
suppressing the present-day \(p\)-wave rate. A dedicated analysis of Fermi-LAT
gamma-ray data gives the representative bound
\(b_{b\bar b}^{\rm lim}<2.4\times10^{-21}~{\rm cm^3\,s^{-1}}\) at
\(m_{\mathcal W}=10~{\rm GeV}\). This bound is roughly ten orders of magnitude
larger than the coefficient given above and therefore does not provide a competitive
constraint on the parameter space considered here
~\cite{KosticLargeScaleStructure2026}. The small dark-matter velocities at
recombination further suppress the \(p\)-wave annihilation rate, so CMB
observations do not provide a relevant constraint in the parameter region considered here~\cite{DiamantiPwaveCMB2014}.

An illustrative thermal average for annihilation into SM fermions gives
$\langle\sigma v_M\rangle\sim3\times10^{-30}~\mathrm{cm^3\,s^{-1}}$
at $m_{\mathcal W}=100~\mathrm{GeV}$, $M_{\mathcal A}=1.67~\mathrm{TeV}$,
$g_4=1$, and $x\equiv m_{\mathcal W}/T=25$.
This suggests additional annihilation channels or nonthermal production, including
the gravitational scenarios developed within this framework
~\cite{TangWuPureGravDM2016,TangWuWeylDM2020,Tang:2020ovf}. A complete
discussion of relic density is beyond the scope of this work.

\subsection{Collider constraints}
\label{subsec:collider}

Collider searches probe virtual mediator exchange in SM fermion-pair production, as well as missing-energy signals from dark-matter production.
Using the spin current in Eq.~\eqref{eq:spin-current} and the identity in
Eq.~\eqref{eq:gamma-identity}, we integrate out the mediator between two
distinct SM fermion currents and obtain the axial--axial contact interaction
used in the LEP-2 comparison,
\begin{equation}
  \Lag_{4f}
  =
  \frac{g_4^2}{8M_{\mathcal A}^2}
  \left(\bar\psi_i\gamma^\lambda\gamma_5\psi_i\right)
  \left(\bar\psi_j\gamma_\lambda\gamma_5\psi_j\right),
  \qquad
  i\neq j,
  \label{eq:collider-contact-operator}
\end{equation}
Matching its coefficient to the conventional $4\pi/\Lambda^2$
normalization gives
$M_{\mathcal A}/g_4=\Lambda/\sqrt{32\pi}\simeq\Lambda/10$.
Since the spin current is flavor universal, the mediator generates
axial--axial contact interactions in both leptonic and hadronic channels.
For the axial--axial contact interaction considered here, we use the LEP-2
limit $\Lambda=16.7~{\rm TeV}$~\cite{LEPEWWG2013}, which yields
\begin{equation}
  \frac{M_{\mathcal A}}{g_4}\gtrsim1.67~{\rm TeV}.
  \label{eq:lep-contact-bound}
\end{equation}
This bound is independent of $m_{\mathcal W}$. For $g_4=1$, the same
coefficient relation gives $M_{\mathcal A}\gtrsim1.67~{\rm TeV}$. Since the LEP-2
$s$-channel satisfies $\sqrt{s}\leq0.209~{\rm TeV}$, the expansion
parameter at the boundary is
$s/M_{\mathcal A}^2\lesssim(0.209/1.67)^2\simeq1.6\times10^{-2}$,
so the contact-interaction approximation is reliable for this choice.

For comparison, the ATLAS dilepton and CMS dijet analyses also report
contact-interaction limits at multi-TeV scales~\cite{ATLASDileptonContact2020, CMSDijetContact2026}. These limits constrain
the coefficient of a local contact operator, while the corresponding
mediator expansion is controlled by the ratio
$|q^2|/M_{\mathcal A}^2=Q^2/M_{\mathcal A}^2$, where
$Q\equiv\sqrt{|q^2|}$ denotes the momentum-transfer scale~\cite{BusoniEFTValidity2014, deVriesFourQuarkEFT2015, RaccoRobustEFT2015}. In the high-energy
bins of the LHC analyses, $Q$ can reach the TeV scale. For $g_4=1$,
the CMS contact limit $\Lambda_{\rm CMS}>19~{\rm TeV}$ gives
$M_{\mathcal A}\simeq1.90~{\rm TeV}$ at the limit when the operator
coefficients are equated. The same consideration applies
to the ATLAS dilepton limit. Consequently, the LHC contact limits do not by
themselves guarantee $Q^2/M_{\mathcal A}^2\ll1$ over the full kinematic
range used in the fits. Accordingly, we quote the LEP-2 result as the formal
collider bound in this work.

\begin{figure}[tb]
  \centering
  \IfFileExists{figures/direct_detection_all_experiments_envelope_with_monojet.pdf}{%
    \includegraphics[width=0.9\textwidth]{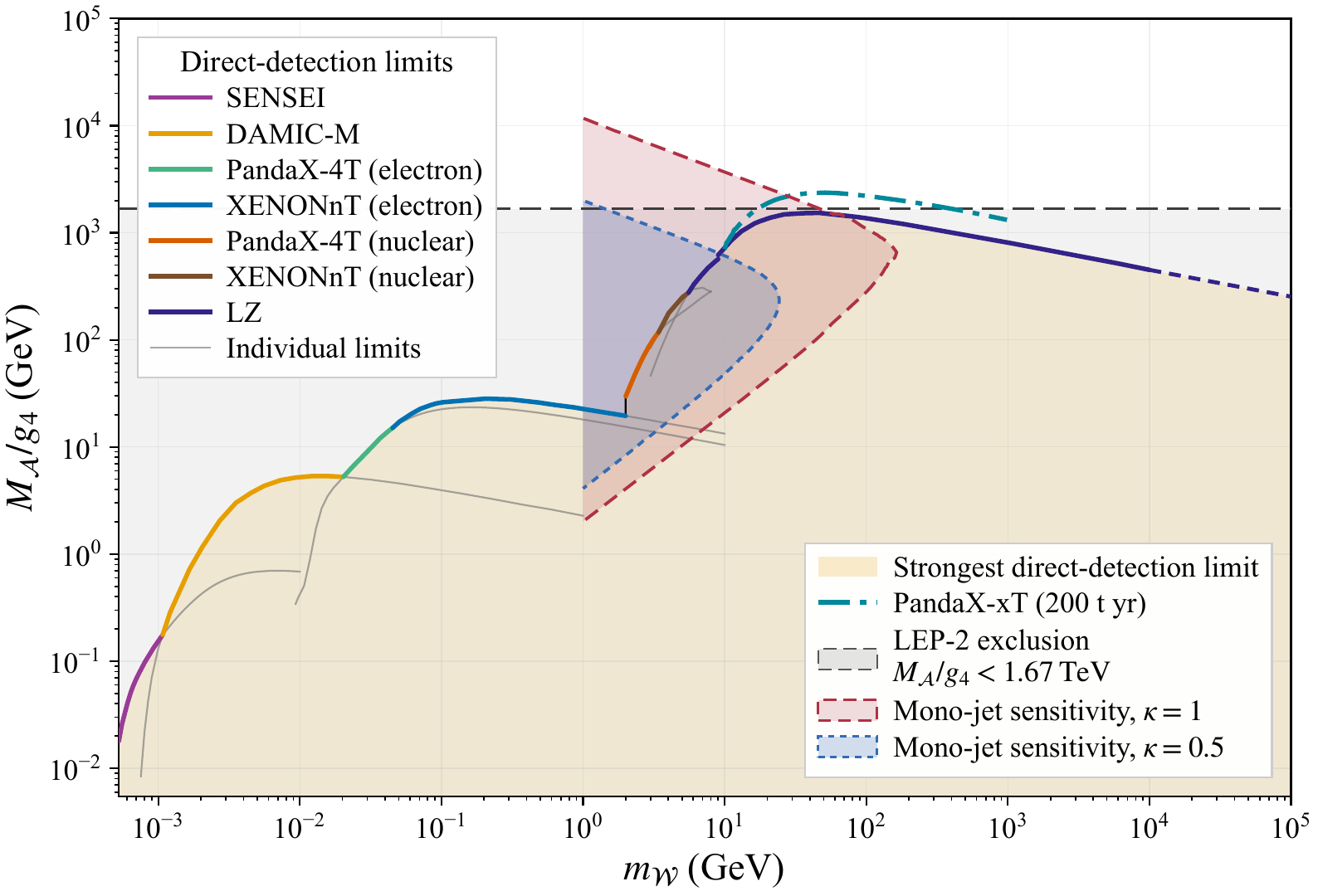}%
  }{%
  \IfFileExists{figures/direct_detection_all_experiments_envelope.pdf}{%
    \includegraphics[width=0.9\textwidth]{figures/direct_detection_all_experiments_envelope.pdf}%
  }{%
    \fbox{\parbox[c][0.35\textwidth][c]{0.70\textwidth}{\centering
    Direct-detection limits and mono-jet sensitivity}}%
  }}
  \caption{\label{fig:direct-detection-envelope}
  Direct-detection limits, the PandaX-xT projection ($200~{\rm tonne\,yr}$),
  the LEP-2 bound, and illustrative mono-jet sensitivities at $g_4=1$.
  Gray curves show individual limits; colored segments mark the strongest
  limit at each mass. Yellow shading extends below this limit and its
  illustrative dashed LZ extrapolation above $10^4~{\rm GeV}$, using the
  high-mass approximation $M_{\mathcal A}/g_4\propto m_{\mathcal W}^{-1/4}$.}
\end{figure}

Jets plus missing transverse momentum provide an established probe of dark matter
at the LHC~\cite{ATLASMETJetsDifferential2024,LHCDarkMatterWG2016}.
In this model, the leading mono-jet processes are
\begin{equation}
  q\bar q\to g\,\mathcal W\mathcal W,
  \qquad
  qg\to q\,\mathcal W\mathcal W,
  \qquad
  \bar qg\to\bar q\,\mathcal W\mathcal W,
  \label{eq:monojet-processes}
\end{equation}
where the visible jet originates from initial-state QCD radiation and the
dark-matter pair is produced through an off-shell
$\mathcal A_\mu^{ab}$. Longitudinal modes enhance mono-jet production for light dark matter.
We compare the leading-order signal rate, evaluated with NNPDF3.1 parton
distributions~\cite{NNPDF31}, with the observed ATLAS upper limits on an
additional signal~\cite{ATLASMonoJet2021}, using the event
selection with the strongest expected sensitivity at each parameter point.
For $g_4=1$, we retain events satisfying
\begin{equation}
  m_{\mathcal W\mathcal W}<
  \min\!\left(M_{\mathcal A},\kappa\Lambda_L\right),
  \qquad
  \Lambda_L\equiv
  \left[4m_{\mathcal W}
  \left(\frac{M_{\mathcal A}}{g_4}\right)^2\right]^{1/3},
  \label{eq:monojet-cutoff}
\end{equation}
where $m_{\mathcal W\mathcal W}$ is the dark-pair invariant mass and
$\Lambda_L$ estimates the validity scale of the effective description
from the energy growth of longitudinal-mode amplitudes.
The dimensionless factor $\kappa$ varies this energy limit; we take
$\kappa=1$ and $0.5$ to illustrate its effect on the sensitivity.
The invariant-mass cutoff reduces the signal and produces the lower
boundaries of the illustrative sensitivity regions.

The mono-photon and mono-$Z$ processes,
$q\bar q\to\gamma\,\mathcal W\mathcal W$ and
$q\bar q\to Z\,\mathcal W\mathcal W$, provide complementary electroweak
signatures. Their production rates are expected to be lower than that of
mono-jet because electroweak radiation is weaker than QCD radiation, while
the leptonic mono-$Z$ channel is further reduced by the branching fraction
and phase space. We therefore focus on mono-jet as the representative LHC missing-energy channel.

\begin{figure}[t]
  \centering
  \IfFileExists{figures/direct_detection_constraints.pdf}{%
    \includegraphics[width=0.9\textwidth]{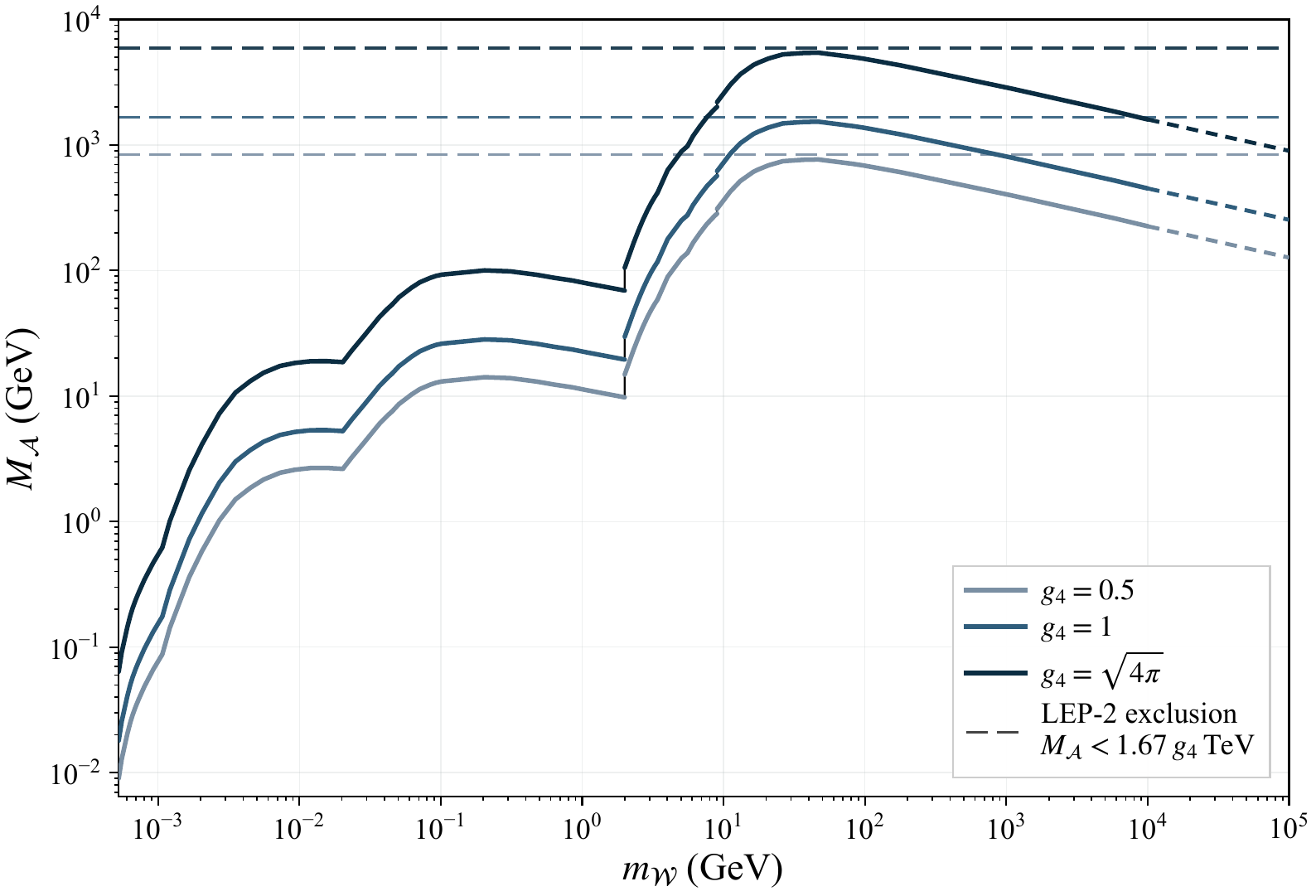}%
  }{%
    \fbox{\parbox[c][0.35\textwidth][c]{0.70\textwidth}{\centering
    Direct-detection constraints in the
    $M_{\mathcal A}$--$m_{\mathcal W}$ plane}}%
  }
  \caption{\label{fig:direct-detection-fixed-g4}
  Direct-detection and LEP-2 constraints for $g_4=0.5$, $1$, and $\sqrt{4\pi}$.
  Regions below the solid curves and horizontal long-dashed lines are
  excluded by direct detection and LEP-2, respectively. Short-dashed tails
  show the illustrative LZ extrapolation of Fig.~\ref{fig:direct-detection-envelope}.}
\end{figure}


Taking all searches together, we can see that current direct-detection searches give their strongest constraint near
$m_{\mathcal W}\simeq50~{\rm GeV}$, requiring
$M_{\mathcal A}/g_4\gtrsim1.54~{\rm TeV}$, slightly below the LEP-2 bound
$M_{\mathcal A}/g_4\gtrsim1.67~{\rm TeV}$. The expected
$200~{\rm tonne\,yr}$ PandaX-xT sensitivity reaches
$M_{\mathcal A}/g_4\simeq2.4~{\rm TeV}$ near the same mass and surpasses
the LEP-2 bound over approximately $m_{\mathcal W}=17$--$400~{\rm GeV}$.
Electron scattering dominates at the low masses. Mono-jet searches provide
additional sensitivity in some mass range.
Due to the velocity suppression, signals from indirect detection are much weaker.

\FloatBarrier

\section{Conclusions}
\label{sec:conclusions}

In this work we developed a low-energy dark-matter model within the GSM: a
stable massive vector $\mathcal W_\mu^a$ coupled to Standard Model fermions
through the heavy spin-gauge mediator $\mathcal A_\mu^{ab}$. Integrating out
the mediator generates a dimension-six axial operator governed by
$M_{\mathcal A}/g_4$. Its Proca form, $Z_2^W$ stability, and
spin-dependent coupling reflect the internal gauge structure.

We have further discussed the relevant phenomenology in direct detection,
indirect detection and collider searches for the vector DM, and illustrated
the constraints and viable parameter space in
Fig.~\ref{fig:direct-detection-envelope}.
Current searches probe $M_{\mathcal A}/g_4$ at the TeV scale.
With a projected exposure of $200~{\rm tonne\,yr}$, PandaX-xT could
probe $M_{\mathcal A}/g_4$ up to about $2.4~{\rm TeV}$ and surpass
the present LEP-2 bound over a broad range of dark-matter masses.

\begin{acknowledgments}
QXZ thanks Guang-Shang Chen and Ziyang Huang for useful discussions. This work is partly supported by the National Key Research and Development Program of China (Grant No.~2021YFC2201901), the National Natural Science Foundation of China (NSFC) under Grants Nos. 12547104, 12441504, 12147103, the Strategic Priority Research Program of the Chinese Academy of Sciences and the Fundamental Research Funds for the Central Universities. 
\end{acknowledgments}

\appendix

\section{Dark matter stability, physical states, and frame conventions}
\label{app:model-construction}
\label{app:frame-direct-detection}

The GQFT framework and GSM construction underlying Sec.~\ref{sec:model}, including the chiral-duality multiplets and their Clifford algebra, are developed in
Refs.~\cite{WuGQFT2016,WuGravidynamics2023,WuGravitizationZeroEMT2025,WuGSMFoundation2025,WuGSMDarkSide2025}.
Here we collect the three ingredients essential for the phenomenological analysis: the symmetry that ensures the stability of $\mathcal W_\mu^a$, the projection onto its physical internal states, and the associated frame conventions.

Let $\Sigma^{ab}$ denote the generators of the intrinsic spin gauge symmetry SP(1,3), and $\Sigma_s^a$ ($s=\mp$) the generators of chirality boost-spin gauge symmetry $W^{1,3}$. The  relevant algebraic relations are
\begin{equation}
  [\Sigma^{ab},\Sigma_s^c]
  =
  i(\Sigma_s^a\eta^{bc}-\Sigma_s^b\eta^{ac}),
  \qquad
  \Sigma_s^a\Sigma_s^b=0 .
  \label{eq:spin-boost-algebra}
\end{equation}
The commutator shows that $\Sigma_s^a$ transforms as an internal vector under
$\mathrm{SP}(1,3)$, providing the translation-like part of the Poincar\'e
analogy discussed in Sec.~\ref{sec:model}.  The vanishing product follows from
the chiral projectors and implies $[\Sigma_s^a,\Sigma_s^b]=0$.  The two values $s=\mp$ label chiral-dual matrix realizations of the same four abstract generators rather than independent copies of $W^{1,3}$, that is, the two chiral-dual generators correspond to the same chirality boost-spin gauge field $\mathcal{W}_{\mu}^a$. 

The chiral-duality equivalence in the GSM induces a $Z_2$ parity in the 16-dimensional spinor representation of chiral leptons and quarks, indicating that the physical laws are independent of the choice of chiral representation. Under this $Z_2$ parity, the boost-spin generators $\Sigma_s^a$ of $W^{1,3}$ flip their chirality, i.e., $\Sigma_-^a\leftrightarrow \Sigma_+^a$. A single-$\mathcal W_\mu^a$ insertion in the chiral fermion kinetic term is proportional to $\Sigma_s^c\Sigma_s^a$ and therefore vanishes. The remaining coupling to the mediator is described by $J_{\mathcal W}$, which is quadratic in $\mathcal W_\mu^a$ [Eq.~\eqref{eq:w-current}].  The retained low-energy action is therefore invariant under
\begin{equation}
  Z_2^{W}:\qquad
  \mathcal W_\mu^a\to -\mathcal W_\mu^a,
  \qquad
  \mathcal A_\mu^{ab}\to \mathcal A_\mu^{ab},
  \qquad
  \mathrm{SM}\to \mathrm{SM},
  \label{eq:appendix-z2}
\end{equation}
Consequently, $\mathcal W_\mu^a$ becomes stable; its interaction with SM fermions is mediated by $\mathcal A_\mu^{ab}$.

As the internal spin gauge symmetry $\mathrm{SP}(1,3)$ is noncompact, the associated covariant tensor metric $\eta_{ab}$ in the spin-linked vector spacetime is not positive definite and thus cannot serve as a summation measure over physical internal states of relevant quantum gauge fields. To that end, it is useful to treat the gauge-field component residing in the coset $\mathrm{SP}(1,3)/\mathrm{SP}(3)$ of the spin gauge group $\mathrm{SP}(1,3)$ as a pure gauge within our low-energy prescription, motivated by constructions with timelike internal directions and dynamical internal metrics~\cite{GielenWise2012,DellDeLyraSmolin1986}. With $S\equiv S(\Lambda)$, we write
\begin{equation}
\begin{aligned}
\mathcal A_\mu\equiv\mathcal A_\mu^{ab}\frac{\Sigma_{ab}}{2}
 &= S\mathcal A_\mu^{ij}\frac{\Sigma_{ij}}{2}S^{-1}
 +\frac{i}{g_4}S\partial_\mu S^{-1},\\
\mathcal W_\mu\equiv\mathcal W_\mu^a\Sigma_{as}
 &= S\mathcal W_\mu^i\Sigma_{is}S^{-1}.
\end{aligned}
\end{equation}
Here $a,b,c,d=0,1,2,3$ and $i,j=1,2,3$.
The fields $\mathcal A_\mu^{ij}$ and $\mathcal W_\mu^i$ on the right-hand sides denote the retained compact-sector variables.
Here $S(\Lambda)$ denotes an internal boost generated by $\Sigma_{0i}$, with conventions
\begin{equation}
\begin{alignedat}{2}
S^{-1}\gamma^a S &= \Lambda^a{}_{b}\gamma^b,
 &\qquad& \Lambda\in\mathrm{SO}^{+}(1,3),\\
S(\Lambda) &= \exp(i\varpi^{0i}\Sigma_{0i}),
 && \varpi^{0i}\in\mathbb R.
\end{alignedat}
\end{equation}
The matrix $\Lambda^a{}_{b}$ is the corresponding Lorentz-type spin boost in the vector representation. Under this prescription, the gauge-field components take the forms
\begin{equation}
\begin{aligned}
\mathcal A_\mu^{ab}
 &=\Lambda^a{}_{i}\mathcal A_\mu^{ij}\Lambda^b{}_{j}
 +\frac{1}{g_4}\Lambda^a{}_{c}\partial_\mu(\Lambda^{-1})^c{}_{d}\eta^{db},\\
\mathcal W_\mu^a
 &=\Lambda^a{}_{i}\mathcal W_\mu^i.
\end{aligned}
\end{equation}

For the physical-state sums in the chosen internal basis, we introduce the unit timelike
internal vector \(n^a=\delta^a_0\), which confines the physical internal components of
\(\mathcal W_\mu^a\) to the three-dimensional subspace orthogonal to
\(n^a\). This in turn allows us to define a covariant tensor $h^{ab}$, in place of the usual metric tensor $\eta^{ab}$, as follows: 
\begin{equation}
h^{ab}
  \equiv
  -\eta^{ab}+n^a n^{b} 
  \label{eq:appendix-internal-physical}; \quad n_a\mathcal W_\mu^a=0.
  \qquad
\end{equation}
Here the tensor \(h^{ab}\) is the internal analog of the massive-vector polarization sum in Minkowski spacetime: it is positive definite on this subspace and implements
the sum over the three physical internal states.
Equation~\eqref{eq:relevant-lagrangian} is written in the full internal-covariant notation prior to this compact-state projection.  In the
physical sector, the quadratic action is contracted with
$h_{ab}=-\eta_{ab}+n_a n_{b}$:
\begin{equation}
  \Lag_{\mathcal W}^{\rm phys}
  =
  -\frac14 h_{ab}\,
  \mathcal W_{\mu\nu}^{a}\mathcal W^{\mu\nu b}
  +
  \frac12m_{\mathcal W}^2 h_{ab}\,
  \mathcal W_\mu^{a}\mathcal W^{\mu b},
  \qquad
  n_a\mathcal W_\mu^a=0 .
  \label{eq:appendix-physical-lagrangian}
\end{equation}
In the orthogonal frame, where $h_{ab}=\operatorname{diag}(0,1,1,1)$,
Eq.~\eqref{eq:appendix-physical-lagrangian} yields the standard Proca quadratic action for each of the three internal components. The dual gravigauge field maps $n^a$ to the timelike spacetime vector $n^\mu=\hat\chi_a{}^\mu n^a=\hat\chi_0{}^\mu$, which reduces to a unit timelike vector in the flat-background limit $\hat\chi_a{}^\mu \to \delta_{a}^{\;\, \mu}$; this vector then provides the basis for constructing the contractions below. 
Throughout the phenomenological analysis, we align this direction
with the cosmological comoving frame, identified at late times
with the CMB rest frame.
The same physical direction is used throughout, with its components
transformed between frames.

The same physical-state projection eliminates the boost-type mediator
components $\mathcal A_\mu^{0i}$ and retains the compact components
$\mathcal A_\mu^{ij}$. The resulting internal contraction is
\begin{equation}
  \sum_{i,j=1}^{3}
  \epsilon^{c ij}{}_{\lambda}\epsilon_{cij\sigma}
  =
  -2\eta_{\lambda\sigma}-4n_\lambda n_\sigma .
  \label{eq:contact-epsilon-contraction}
\end{equation}

Writing
$j_{5,f}^{\lambda}\equiv\bar\psi_f\gamma^\lambda\gamma_5\psi_f$ and
$J_5^\lambda\equiv\sum_f j_{5,f}^\lambda$, integrating out the mediator gives
\begin{equation}
  \Lag_{4f}
  =
  \frac{g_4^2}{16M_{\mathcal A}^2}
  \left[
    J_5^\lambda J_{5\lambda}
    +2(n_\lambda J_5^\lambda)^2
  \right].
  \label{eq:appendix-contact-operator}
\end{equation}
Upon expanding the total current, the coefficients of the distinct- and
same-flavor terms are $g_4^2/(8M_{\mathcal A}^2)$ and
$g_4^2/(16M_{\mathcal A}^2)$, respectively. The latter matches the CMS
$2\pi/\Lambda^2$ convention for the Lorentz-invariant term and gives the
same relation $M_{\mathcal A}/g_4=\Lambda/\sqrt{32\pi}$
~\cite{CMSDijetContact2026}.

For the universal axial interaction considered here, the
$s$-channel process $e^+e^-\to\mu^+\mu^-$ provides the most direct
LEP-2 constraint.
Neglecting the laboratory velocity relative to the cosmological frame,
$q_s^\mu\equiv p_{e^-}^\mu+p_{e^+}^\mu=(\sqrt{s},\mathbf 0)$ and
$n^\mu\simeq q_s^\mu/\sqrt{s}$, so the additional $n^\mu n^\nu$ term is
Ward-suppressed by $m_e m_\mu/s$. This argument does not apply to the
$t$-channel in Bhabha scattering, whose momentum transfer is spacelike
and angle dependent; we therefore do not use the Bhabha-inclusive
combination.

The projector $h^{aa'}$ in Eq.~\eqref{eq:appendix-internal-physical}
implements the sum over the physical internal states of
quantum gauge field $\mathcal W_\mu^a$ in Eq.~\eqref{eq:scattering-amplitude}. This
internal-state sum then introduces the following contraction tensor
\begin{equation}
  \mathcal H^{\nu\rho;\nu'\rho'}
  =
  h^{aa'}h^{bb'}
  \hat\chi_a{}^\nu\hat\chi_b{}^\rho
  \hat\chi_{a'}{}^{\nu'}\hat\chi_{b'}{}^{\rho'} .
  \label{eq:appendix-h-tensor}
\end{equation}
For the external states we use the massive-vector polarization sum
\begin{equation}
  \sum_\zeta
  \varepsilon_\alpha(k,\zeta)
  \varepsilon_\gamma^*(k,\zeta)
  =
  \Pi_{\alpha\gamma}(k)
  =
  -\eta_{\alpha\gamma}
  +
  \frac{k_\alpha k_\gamma}{m_{\mathcal W}^2},
  \label{eq:vector-polarization-sum}
\end{equation}
and the axial fermion spin trace for elastic scattering is
\begin{align}
  S^{\lambda\lambda'}
  &=
  \mathrm{Tr}\left[
    (\gamma\cdot p_2+m_f)\gamma^\lambda\gamma_5
    (\gamma\cdot p_1+m_f)\gamma^{\lambda'}\gamma_5
  \right]
  \nonumber\\
  &=
  4\left[
    p_2^\lambda p_1^{\lambda'}
    +
    p_2^{\lambda'}p_1^\lambda
    -
    (p_1\cdot p_2+m_f^2)\eta^{\lambda\lambda'}
  \right].
  \label{eq:axial-spin-trace}
\end{align}
Here $\Pi_1=\Pi(k_1)$ and $\Pi_2=\Pi(k_2)$. Before averaging over the
initial states, the squared amplitude is therefore
\begin{align}
  |\mathcal M|^2
  &=
  \frac{g_4^4}{16M_{\mathcal A}^4}
  \widetilde V^{\mu\alpha\beta}
  \widetilde V^{\mu'\gamma\delta}
  \Pi_{1,\alpha\gamma}
  \Pi_{2,\beta\delta}
  \epsilon_{\mu\nu\rho\lambda}
  \epsilon_{\mu'\nu'\rho'\lambda'}
  \mathcal H^{\nu\rho;\nu'\rho'}
  S^{\lambda\lambda'} .
  \label{eq:appendix-frame-squared}
\end{align}
The initial-state averaging factors are those of
Sec.~\ref{subsec:direct-detection}.

The gravigauge field $\hat\chi_a{}^\mu$ generally enters through the symmetric combination
\begin{equation}
  P_\chi^{\mu\nu}
  =
  h^{ab}\hat\chi_a{}^\mu\hat\chi_{b}{}^\nu .
  \label{eq:appendix-pchi}
\end{equation}
In a flat background, a frame related to the fiducial one by
a Lorentz-type spin boost $\Lambda(\varpi)$ may be parametrized as
\begin{equation}
  \chi_\mu{}^a
  =
  \Lambda^{a}_{\;\; b}(\varpi)\,\delta_\mu{}^b,
  \qquad
  \Lambda^T\eta \Lambda=\eta ,
  \label{eq:appendix-boost-frame}
\end{equation}
so that
\begin{equation}
  P_\chi^{\mu\nu}
  =
  -\eta^{\mu\nu}
  +
  n^\mu n^\nu,
  \label{eq:appendix-pchi-n}
\end{equation}
and Eq.~\eqref{eq:appendix-h-tensor} becomes
\begin{equation}
  \mathcal H^{\nu\rho;\nu'\rho'}
  =
  P_\chi^{\nu\nu'}P_\chi^{\rho\rho'} .
  \label{eq:appendix-h-pchi}
\end{equation}
Thus Eq.~\eqref{eq:appendix-frame-squared} can be collected into the covariant tensor
\begin{align}
  \mathcal K_{\mu\mu'\lambda\lambda'}
  &\equiv
  \epsilon_{\mu\nu\rho\lambda}
  \epsilon_{\mu'\nu'\rho'\lambda'}
  P_\chi^{\nu\nu'}P_\chi^{\rho\rho'}
  \nonumber\\
  &=
  2\left(
    n_\mu n_{\mu'}P_{\chi\,\lambda\lambda'}
    -n_\mu n_{\lambda'}P_{\chi\,\lambda\mu'}
    -n_\lambda n_{\mu'}P_{\chi\,\mu\lambda'}
    +n_\lambda n_{\lambda'}P_{\chi\,\mu\mu'}
  \right),
  \label{eq:appendix-k-tensor}
\end{align}
where indices on $P_\chi^{\mu\nu}$ are lowered with $\eta_{\mu\nu}$.  The
covariant squared amplitude for elastic scattering can therefore be written as
\begin{align}
  |\mathcal M|^2_{\mathrm{cov}}
  &=
  \frac{g_4^4}{16M_{\mathcal A}^4}
  \widetilde V^{\mu\alpha\beta}
  \widetilde V^{\mu'\gamma\delta}
  \Pi_{1,\alpha\gamma}
  \Pi_{2,\beta\delta}
  \mathcal K_{\mu\mu'\lambda\lambda'}
  S^{\lambda\lambda'} .
  \label{eq:appendix-covariant-squared}
\end{align}
This is Eq.~\eqref{eq:appendix-frame-squared} with all frame-dependence
encoded in the unit timelike direction \(n^\mu\) selected by the spin gauge prescription.  
The laboratory and halo frames move relative to this cosmological
frame with velocities of order $10^{-3}$.
These small boosts do not modify the leading velocity-independent
terms retained in the direct-detection analysis.
For mono-jet production, we likewise neglect the laboratory boost,
while retaining the motion of the outgoing DM pair in the laboratory.

For annihilation, Eq.~\eqref{eq:annihilation-pwave-cross-section}
applies when the incoming DM pair is at rest in the cosmological
frame, so that $n^\mu=q^\mu/\sqrt{s}$, where $q=k_1+k_2$ and $s=q^2$.
For a moving pair, its squared center-of-mass speed in this frame is
$u^2=1-s/(n\cdot q)^2$.
At small $u$, pair motion introduces a helicity-unsuppressed
correction of order $u^2v_{\rm rel}^2$; the rate remains $p$-wave
suppressed at fixed $u$ and negligible for the cold-halo benchmarks.
For the thermal estimate in Sec.~\ref{subsec:indirect-detection},
we take the DM bath to be at rest in the same cosmological frame
and average the two incoming velocities independently over a
nonrelativistic Maxwell--Boltzmann distribution.
This gives $\langle v_{\rm rel}^2\rangle=6/x$ and
$\langle u^2v_{\rm rel}^2\rangle=9/x^2$, with $x=m_{\mathcal W}/T$.
With equal weights for the initial physical polarizations and
internal states, the leading M\o ller rate for
$m_f\ll m_{\mathcal W}$ away from thresholds is
\begin{equation}
 \langle\sigma v_M\rangle\simeq
 \frac{7}{2592\pi R^4}\sum_{f\,\mathrm{open}}N_c^f
 \left(\frac{6m_f^2}{x}+\frac{8m_{\mathcal W}^2}{x^2}\right),
 \qquad R\equiv\frac{M_{\mathcal A}}{g_4}.
 \label{eq:illustrative-thermal-rate}
\end{equation}
The second term accounts for the center-of-mass motion of the
annihilating pairs.
The benchmark sums over open charged SM channels at an illustrative $x=25$,
without determining the freeze-out temperature.

\interlinepenalty=10000 
\bibliography{references}

\end{document}